# Evaluating the Effectiveness of Regional Lockdown Policies in the Containment of Covid-19: Evidence from Pakistan

**Hamza Umer**
Ph.D. Economics
Graduate School of Economics
Waseda University, Tokyo Japan
Email: hamzaumer@akane.waseda.jp

**Muhammad Salar Khan**
Ph.D. Candidate, Public Policy
Schar School of Policy and Government
George Mason University, Arlington Virginia USA
Email: khan63@gmu.edu

**Abstract**: To slow down the spread of Covid-19, administrative regions within Pakistan imposed *complete* and *partial* lockdown restrictions on socio-economic activities, religious congregations, and human movement. Here we examine the impact of regional lockdown strategies on Covid-19 outcomes. After conducting econometric analyses (*Regression Discontinuity* and *Negative Binomial* Regressions) on official data from the National Institute of Health (NIH) Pakistan, we find that the strategies did not lead to a similar level of Covid-19 caseload (*positive cases* and *deaths*) in all regions. In terms of reduction in the overall caseload (positive cases and deaths), compared to *no lockdown, complete* and *partial lockdown* appeared to be effective in four regions: Balochistan, Gilgit Baltistan (GT), Islamabad Capital Territory (ICT), and Azad Jammu and Kashmir (AJK). Contrarily, complete and partial lockdowns did not appear to be effective in containing the virus in the three largest provinces of Punjab, Sindh, and Khyber Pakhtunkhwa (KPK). The observed regional heterogeneity in the effectiveness of lockdowns advocates for a careful use of lockdown strategies based on the demographic, social, and economic factors.

## 1. Introduction

The world is struggling to combat the Covid-19 pandemic, which has spread from Wuhan China to over 190 countries. Unless there is a viable treatment or vaccine to treat Covid-19, the world is taking possible preventive measures to minimize the spread of the disease.

Lockdown (complete or partial) is one of the most evident and widely used preventive measures. The effectiveness of lockdown in controlling the spread of Covid-19, however, is not a well-established outcome for a few reasons. First, so much is unknown about the novel coronavirus and that the situation is evolving, which gives policymakers little time to think through, implement, and properly investigate or foresee the effectiveness of any policy, such as a lockdown. Second, while many countries—for instance, New Zealand and Germany—have already implemented complete or partial lockdown (McFall-Johnsen et al., 2020), it is just hard to conduct a long-term ex-ante analysis of the repercussions of policy concerning *once-in-a-century* pandemic. Still, some preliminary studies tried to investigate the effect of lockdown strategies on the spread of Covid-19. For instance, Walker and Colleagues (2020) predicted that interventions and lockdown strategies in almost all countries (precisely, 202 countries in their analysis) would reduce infections and deaths by nearly half. Mushfiq Mobarak and Colleagues (2020), contrarily argued that while lockdown strategies are viable options in high-income countries, low-income countries such as Nigeria and Pakistan cannot afford to have a fruitful lockdown. Because of their weak capacity in enforcing lockdown strategies, these countries may witness counterproductive effects if such strategies make workers and migrants migrate back from heavily populated urban areas and spread the disease to remote rural areas, the researchers argue.

Similarly, we have a few studies that show different outcomes on the country-level. For example, research (Dowd et al., 2020) on the Italy outbreak shows the effectiveness of early lockdown. In Italy, the Covid-19 was first detected in the Lodi province, which placed restrictions beginning February 22. As opposed to this, Bergamo province, which started with fewer cases but did not impose restrictions until March 7, far surpassed the number of cases in Lodi (Stancati, 2020). In a similar vein, a study (Kumar and Nataraj, 2020) on Indian regions shows a regional differential in the spread of Covid-19 in the face of regional policy variation.

Overall, the world is divided regarding the use and effectiveness of lockdown policies. On one side, we see countries like Japan (Du and Huang, 2020; Ian, 2020) and Sweden (Karlson et al., 2020) that have used mild lockdown or no lockdown and yet effectively contained the spread of the virus. On the other side, we see countries like New Zealand and Australia using strong lockdown policies to flatten the spread of the virus (Fifield, 2020).

This article offers a systematic contribution to the aforementioned debate by examining the effectiveness of lockdown policies in the containment of Covid-19 in the context of Pakistan. Specifically, the article explores how effective the lockdown strategy has been in combating Covid-19 outcomes in the country. This evaluation of lockdown policies is based on the econometric analysis (such as regression discontinuity and negative binomial regressions) performed on official data from Pakistan. Pakistan is selected because it offers a valuable opportunity to analyze the effects of both complete lockdown and partial lockdown policies on the spread of the Covid-19 virus. Moreover, the regional use of lockdown policies in Pakistan is heterogeneous and hence enables us to perform cross-regional analysis as well. We find that in comparison to no lockdown, complete and partial lockdowns have been ineffective in the containment of the virus in the three largest provinces Punjab, Sindh, and Khyber Pakhtunkhwa (KPK). On the other hand, complete and partial lockdowns have been very effective in the containment of the virus in the province of Balochistan and the three administrative territories/regions of Gilgit Baltistan (GB), Islamabad Capital Territory (ICT), and Azad Jammu and Kashmir (AJK). The observed regional heterogeneity in the effectiveness of lockdowns advocates for a careful use of lockdown based on the demographic, societal, and economic factors. “One size fits all” approach for lockdown could be counterproductive in some regions of the world and subsequently make the spread of virus more acute, as demonstrated by researchers in the context of Africa as well (Mehtar et al., 2020).

## 2. Pakistan – A Brief History of Covid-19

Pakistan—home to about 220 million and wobbly health infrastructure that has close to 1.5 million hospital beds (Khan & Latif, 2020)—reported the first case of the Covid-19 on February 26, a returning pilgrim from neighboring Iran (Hashim, 2020). On the same day, the Pakistan Federal Ministry of Health confirmed another positive case in Islamabad (Ali, 2020). Since then, the virus has diffused quickly. By March 18, all the administrative regions of Pakistan, including four provinces (Punjab, Sindh, KPK, and Balochistan), the two autonomous territories (AJK and GB), and the federal territory of Islamabad registered positive cases. The entire country reported over eighty-five thousand confirmed cases and 1,770 deaths, as of June 4, 2020.[1] In terms of the total number of cases and deaths, Pakistan ranks 17th and 21th worldwide, respectively.[2] While now the virus has entered the community transmission stage, initially, all the confirmed cases in Pakistan had recent travel history from Iran, Syria, London, and Saudi Arabia.

The Covid-19 pandemic has spread unevenly across regions within Pakistan, with the four regions (provinces) making up more than 95 percent of the cases as of June 4, 2020. Sindh registered the most cases at over 39,900, followed by Punjab (31,104), KPK (11,373), and Balochistan (5,224).[3] The province of Punjab reported 607 deaths, the most in the country, followed by Sindh (555) and KKP (500) and Balochistan (51). The situation in the three special regions or territories is not that bad, with Islamabad, GB, and AJK reporting 3,544,

---

[1] This data was obtained from Worldometers: https://www.worldometers.info/coronavirus/

The data was cross-verified here: https://www.cdc.gov/coronavirus/2019-ncov/global-covid-19/world-map.html

[2] Ibid.

[3] Official government website for data on Covid-19 was consulted: http://covid.gov.pk/stats/pakistan

824, and 285 confirmed cases, respectively. The three regions together reported 57 deaths. Overall, the mortality rate in Pakistan is 2%, and the recovery rate is 35%.[4]

**Figure 1: Daily Covid-19 New Cases in Pakistan**

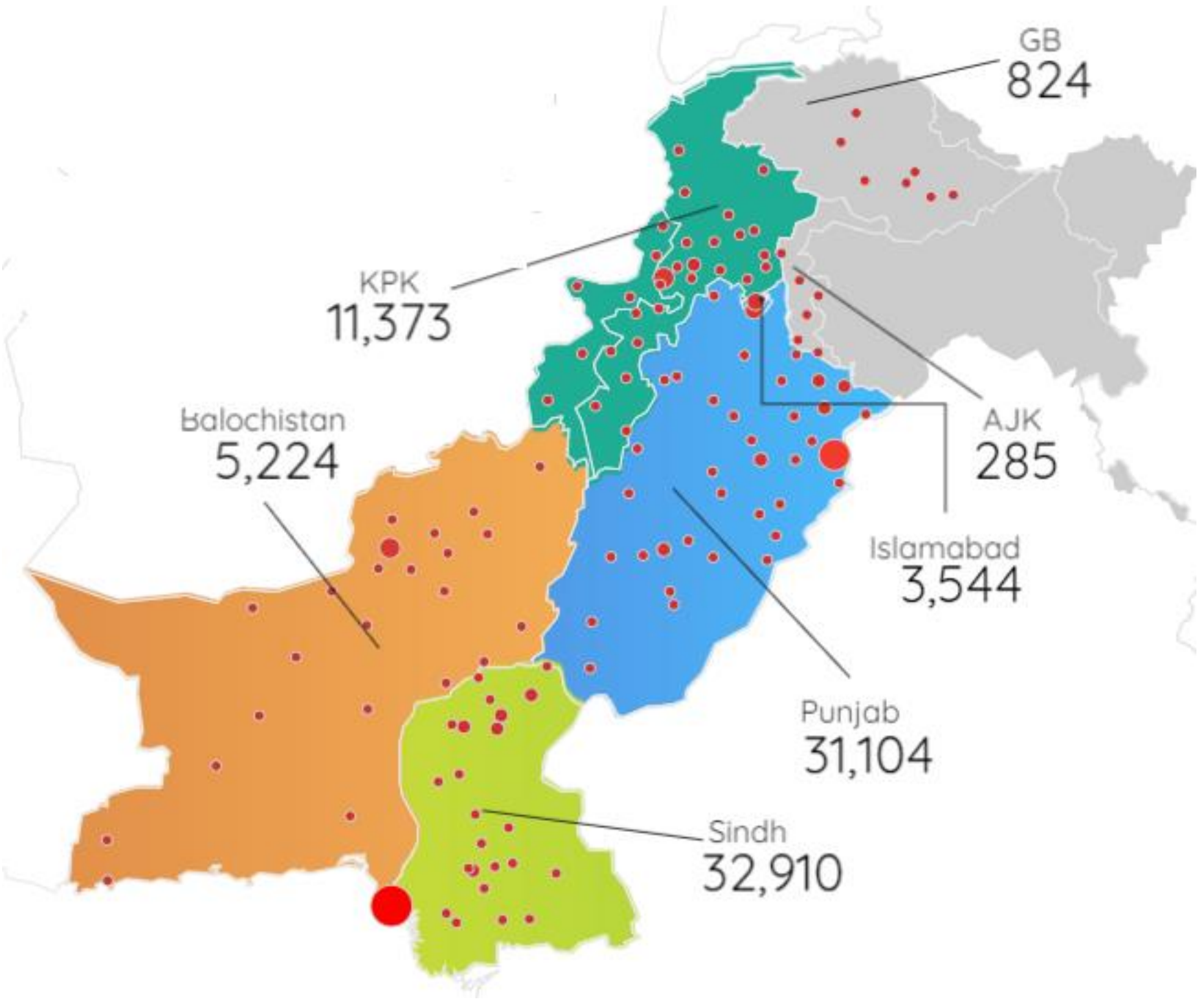

Source: http://covid.gov.pk/stats/pakistan

As a nascent federalist country,[5] when the Interior Ministry of Pakistan announced a lockdown on March 23 to combat the spread of the virus, all the seven administrative regions also implemented their regional lockdown measures at or around March 23. Army troops were deployed throughout the country to help the divisions in tackling the spread of the virus.[6] Initially, the regions implemented a full or complete lockdown, with ICT shutting down as early as on March 12, Punjab on March 22, and all other divisions on March 24.[7]

---

[4] Ibid.

[5] Pakistan is a federalist country, with provincial governments having the right to decide on important issues, according to the Eighteenth Amendment of the Constitution of Pakistan.

[6] https://www.geo.tv/latest/278812-government-calls-in-pakistan-army-troops-amid-coronavirus-outbreak

[7] Complete lockdown and partial lockdown information is extracted from several newspapers and online sources. These include: Business Recorder, Dawn News, Radio Pakistan, Technology Times. Also please refer to the report (Covid-19 Legislation and Measures in Pakistan) published by Zafar Kalanuri (2020). Complete lockdown refers to complete shutdown of socioeconomic, religious activities and mobility pathways while partial lockdown refers to controlled opening of the aforementioned. Complete description of lockdown types is mentioned on page 6.

The duration of this full lockdown also varied, with ICT and KPK observing a short duration lockdown for less than a week, Punjab, Sindh, and Balochistan observing a medium duration lockdown for about two weeks, and AJK and GB observing a long duration lockdown for almost a month. Later, the divisions moved to a partial or controlled lockdown (please see Table 3 for details).

Policy-wise, Pakistan acted quickly and formulated a National Action Plan for Covid-19 early in February 2020 (Mukhtar, 2020). The Ministry of National Health Services, Regulation & Coordination Pakistan presented the Plan, that was supposed "to provide (a) policy framework for federal, provincial, and regional stakeholders for building capacity to prevent, detect and respond to any events due to COVID-2019 in Pakistan."[8] Along with the health ministry, National and Provincial Disaster Management Authorities, National Command and Operation Center (NCOC), and National Coordination Committee (NCC) have been formulating, coordinating, analyzing, and implementing policy efforts about Covid-19. With federal directives, regions have been managing outbreaks according to their circumstances.

Table 1 provides details about the total number of positive cases, deaths caused by Coivd-19 and total tests performed in the seven regions of Pakistan.

**Table 1: Summary of Covid-19 Cases and Deaths**

| Region | Total Positive | Total Deaths | Total Tests |
|---|---|---|---|
| **ICT** | 677 | 6 | 19452 |
| **Punjab** | 11,568 | 197 | 130,451 |
| **Sindh** | 11,466 | 189 | 91,152 |
| **KPK** | 4669 | 245 | 30,444 |
| **Balochistan** | 2016 | 26 | 14,669 |
| **AJK** | 86 | 0 | 2850 |
| **GB** | 440 | 4 | 5454 |
| **Total** | 30,922 | 667 | 294,472 |

Note: The data is from March 12, 2020 till May 11, 2020.

As evident from Table 1, in terms of the number of positive cases and deaths, Punjab, Sindh, and KPK are the worst struck regions by Covid-19. AJK is the least affected region with no deaths.

[8]National Action Plan for Corona virus disease (COVID-19), National Institute of Health Pakistan

## 3. Data Sources & Data Description

We analyze the effect of lockdown policies on two outcome variables (daily positive cases and daily deaths due to Covid-19) for four provinces Punjab, Sindh, KPK, Balochistan, and the three regions of GB, ICT, and AJK. The data is acquired from *Kaggle*,[9] which in turn is based on the information shared by the National Institute of Health Pakistan (NIH).[10] We cross-verified the data with the official data publications listed on the NIH website. The data included cumulative numbers for positive cases of Covid-19 and deaths due to Covid-19. The cumulative positive cases and deaths are transformed into daily positive cases and daily deaths by taking the first difference of the cumulative numbers. Each of the seven regions in Pakistan had its lockdown policies. Based on the strength of their stringency, we classify lockdown into two categories: complete lockdown and partial lockdown. Complete lockdown refers to complete closure of market activities (except necessary items such as food), schools, workplaces, parks and other public places, ban on social gatherings and social events, closing of land and air transport and people are restricted to stay home unless they need medical help or require grocery shopping. Partial lockdown refers to the controlled opening of economic activities for a specific time every day, limited resumption of land and air transport, and maintaining social distance during outdoor activities. Most of the educational institutes, however, remain closed.[11] The data covers a time period of 60 days (March 12 to May 11, 2020) and encompasses no lockdown, complete lockdown, and partial lockdown phases. The entire data variables are described in Table 2.

---

[9] Hemani, Mesum Raza. "Corona Virus Pakistan Dataset 2020." Accessed May 31, 2020. https://kaggle.com/mesumraza/corona-virus-pakistan-dataset-2020.Kaggle—a subsidiary of Google LLC—allows to find and publish data sets, explore and build models in a web-based data-science environment, work with other data scientists and machine learning engineers, and enter competitions to solve data science challenges.
https://www.kaggle.com/

[10] Please see : https://www.nih.org.pk/novel-coranavirus-2019-ncov/

[11] Please see these sources: https://www.pakistantoday.com.pk/2020/04/02/covid-19-economic-impact-nationwide-lockdown/;https://www.dawn.com/news/1545369;https://92newshd.tv/complete-lockdown-imposed-in-ajk-for-three-weeks/#.XshCxxsw_mI;https://www.reuters.com/article/us-health-coronavirus-pakistan-idUSKBN22J1MX;https://www.geo.tv/latest/278812-government-calls-in-pakistan-army-troops-amid-coronavirus-outbreak

**Table 2: Description of Data Variables**

| Variables | Description |
|---|---|
| Daily Deaths | Number of daily deaths due to Covid-19 in Punjab, Sindh, KPK, Balochistan, GB, ICT, and AJK. |
| Daily Positive Cases | Number of daily positive cases of Covid-19 in Punjab, Sindh, KPK, Balochistan, GB, ICT, and AJK. |
| Daily Tests Performed | Number of daily tests of Covid-19 in Punjab, Sindh, KPK, Balochistan, GB, ICT, and AJK. |
| Lockdown | A dummy variable that takes on a value of one from the day complete lockdown is imposed and zero otherwise. As each region in Pakistan imposed a lockdown on separate days, we have seven lockdown variables representing seven regions. |
| Partial Lockdown | A dummy variable that takes on a value of one from the day partial complete lockdown is imposed and zero otherwise. As each unit in Pakistan imposed a partial lockdown on separate days, we have seven lockdown variables representing seven regions. |
| Days | It is a time variable that represents day. |

In Table 3, the summary of the variables is reported. Punjab, Sindh, and KPK are three provinces witnessing the highest number of average daily positive cases and average daily deaths. ICT had the longest complete lockdown (34 days), while KPK had the shortest one (5 days).

**Table 3: Summary of Variables**

| | ICT | Punjab | Sindh | KPK | Balochistan | AJK | GB |
|---|---|---|---|---|---|---|---|
| **Daily Deaths** | n = 60 m = 0.1 (0.3) | n = 60 m = 3.28 (4.54) | n = 60 m = 3.15 (3.62) | n = 60 m = 4.08 (5.06) | n = 59 m = 0.46 (0.92) | n = 60 m = 0* | n = 60 m = .0.07 (0.25) |
| **Daily Positive Cases** | n = 58 m= 11.79 (13.56) | n = 60 m =192.8 (211.79) | n = 59 m=195.41 (233.08) | n = 60 m = 77.82 (88.69) | n = 59 m = 34.19 (42.30) | n = 60 m = 1.43 (2.09) | n = 60 m = 7.33 (7.59) |
| **Daily Tests Performed** | n= 59 m=337.20 | n= 58 m=2266.83 | n=60 m=1519.2 | n= 57 m=544 | n=57 m=258.07 | n=60 m=47.5 | n=58 m=96.84 |

| | ICT | Punjab | Sindh | KPK | Balochistan | AJK | GB |
|---|---|---|---|---|---|---|---|
| | (322.87) | (2482.48) | (1669.13) | (611.71) | (228.87) | (42.49) | (73.80) |
| **Time** | 3/12--5/11 | 3/12--5/11 | 3/12--5/11 | 3/12--5/11 | 3/12--5/11 | 3/12--5/11 | 3/12--5/11 |
| **Lockdown**[12] | 3/12--4/15 | 3/22--4/7 | 3/24--4/6 | 3/24--3/28 | 3/24--4/07 | 3/24--4/24 | 3/24--4/21 |
| **Partial Lockdown** | 4/16--5/11 | 4/8--5/11 | 4/7--5/11 | 3/29--5/11 | 4/08--5/11 | 4/25--5/11 | 4/22--5/11 |

Note: n = Observations; m = Mean. Standard deviations are in parentheses. *AJK did not report any death during the data duration specified in this paper.

## 4. Econometric Analysis

All the econometric analyses are performed in *STATA* 16. The outcome variables include the number of daily deaths due to Ccovid-19 and the number of people testing positive for Ccovid-19. The main explanatory variables are lockdown and partial lockdown dummies.

### 4.1. Binned Scatter Plots & Regression Discontinuity

The starting point of the analysis is a visual representation of the impact of lockdown policies on the outcome variables. This is achieved by using regression discontinuity (RD) with the date as *running* variable and the lockdown and partial lockdown dates as multiple cutoffs. In *STATA* it is implemented with the help of a binned scatter plot technique[13] with complete lockdown and partial lockdown dummies accounting for regression discontinuities (Chetty et al., 2011; Cattaneo et al., 2017). The binned scatter plots for four regions (Punjab, Sindh, Balochistan, and KPK) cumulatively representative of 96% of Covid-19 cases are in Figure 2. As these four provinces account for the majority of the positive cases, we discuss them at length here, while binned scatter plots for ICT, AJK, and GB regions are reported in appendix A.

[12] Complete lockdown and partial lockdown information is extracted from several news papers and online sources. These include: Business Recorder, Dawn News, Radio Pakistan, Technology Times and also please refer to the report (Covid-19 Legislation and Measures in Pakistan) published by Zafar Kalanuri (2020). Please check out these links:
https://www.brecorder.com/2020/04/08/587539/gilgit-baltistan-extends-lockdown-till-april-21/
https://www.dawn.com/news/1550885
https://www.radio.gov.pk/22-03-2020/gb-govt-decides-to-observe-lockdown-for-indefinite-period
https://www.technologytimes.pk/2020/04/08/coronavirus-spread-timeline-pakistan/

**Figure 2: Lockdown Policies & Spread of Covid-19**

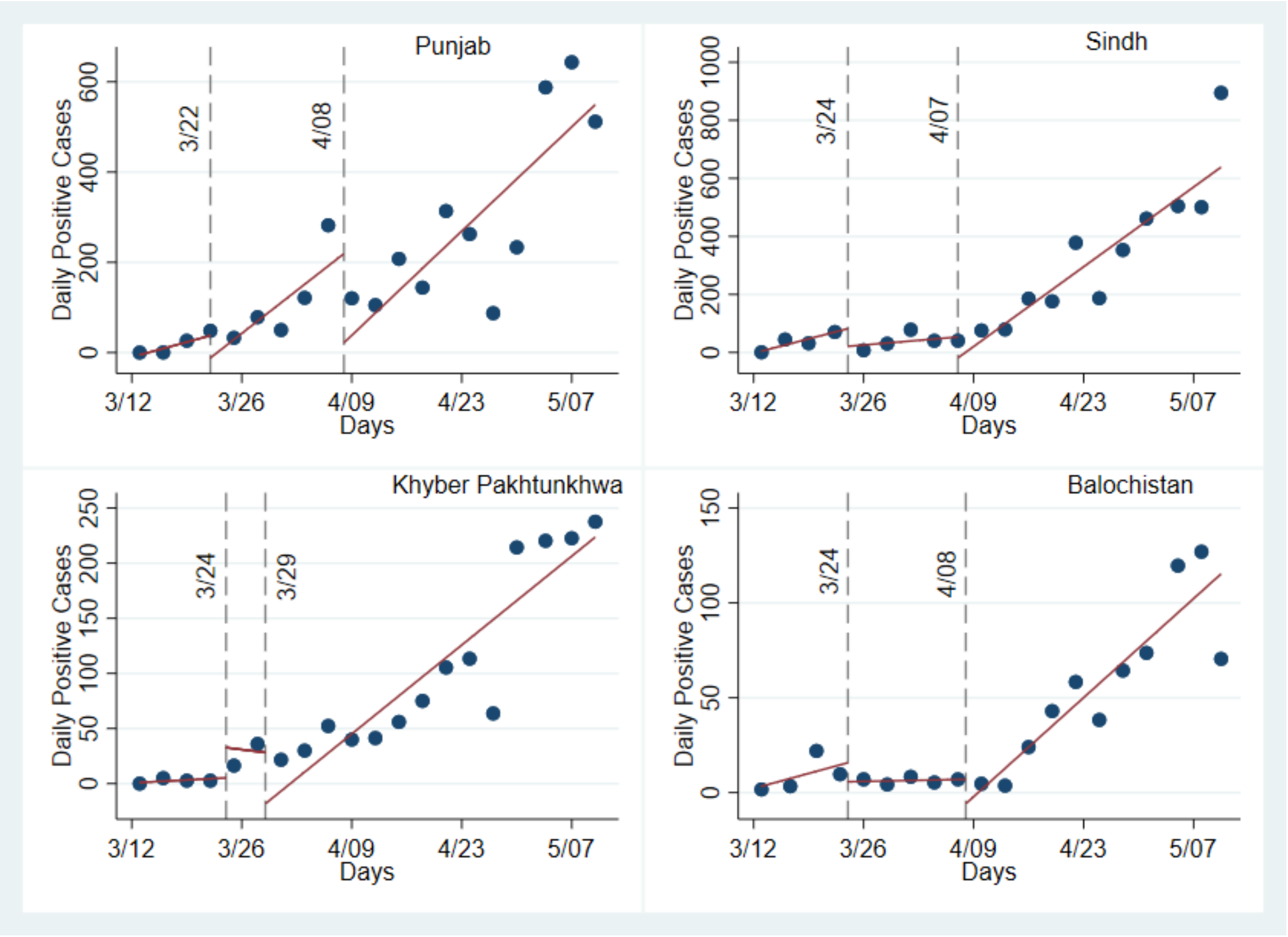

Note: The leftmost vertical dotted line indicates the introduction of a complete lockdown. The rightmost vertical line indicates the introduction of partial lockdown.

The Covid-19 positive cases in Punjab and Sindh decrease following the implementation of complete lockdown. However, positive cases increase in both regions towards the end of complete lockdown. Following the implementation of partial lockdown, the numbers of positive cases initially decrease in Punjab while increasing in Sindh, and over time increase in both regions. Overall we observe an increasing trend in positive cases during the complete and partial lockdown phases (as indicated by the regression lines). On the other hand, positive cases in KPK increase following the implementation of complete lockdown and keep on rising towards the end of lockdown as well. The lockdown duration in KPK, however, is very short (just five days). Following the implementation of partial lockdown, positive cases decrease; however, they start increasing with time. Overall we observe an increasing trend during the lockdown phases. Positive cases in Balochistan decrease following the introduction of lockdown, and they continue to fall towards the end of lockdown as well. However, following the implementation of partial lockdown, the number of positive cases increase. Complete lockdown appears to be effective in controlling the spread of the virus in Balochistan only while partial lockdown is apparently ineffective in all four regions.

[13] The analysis is performed using *binscatter* command and 20 bins (default setting of STATA) for the binned scatter plots.

Next, we analyze the number of daily deaths caused by Covid-19. The binned scatter plots for four regions (Punjab, Sindh, Balochistan, and KPK), accounting for 95% of the deaths in Pakistan, are represented in Figure 3. To conserve space, binned scatter plots for ICT, AJK, and GB regions are reported in appendix B.

**Figure 3: Lockdown Policies & Deaths due to Covid-19**

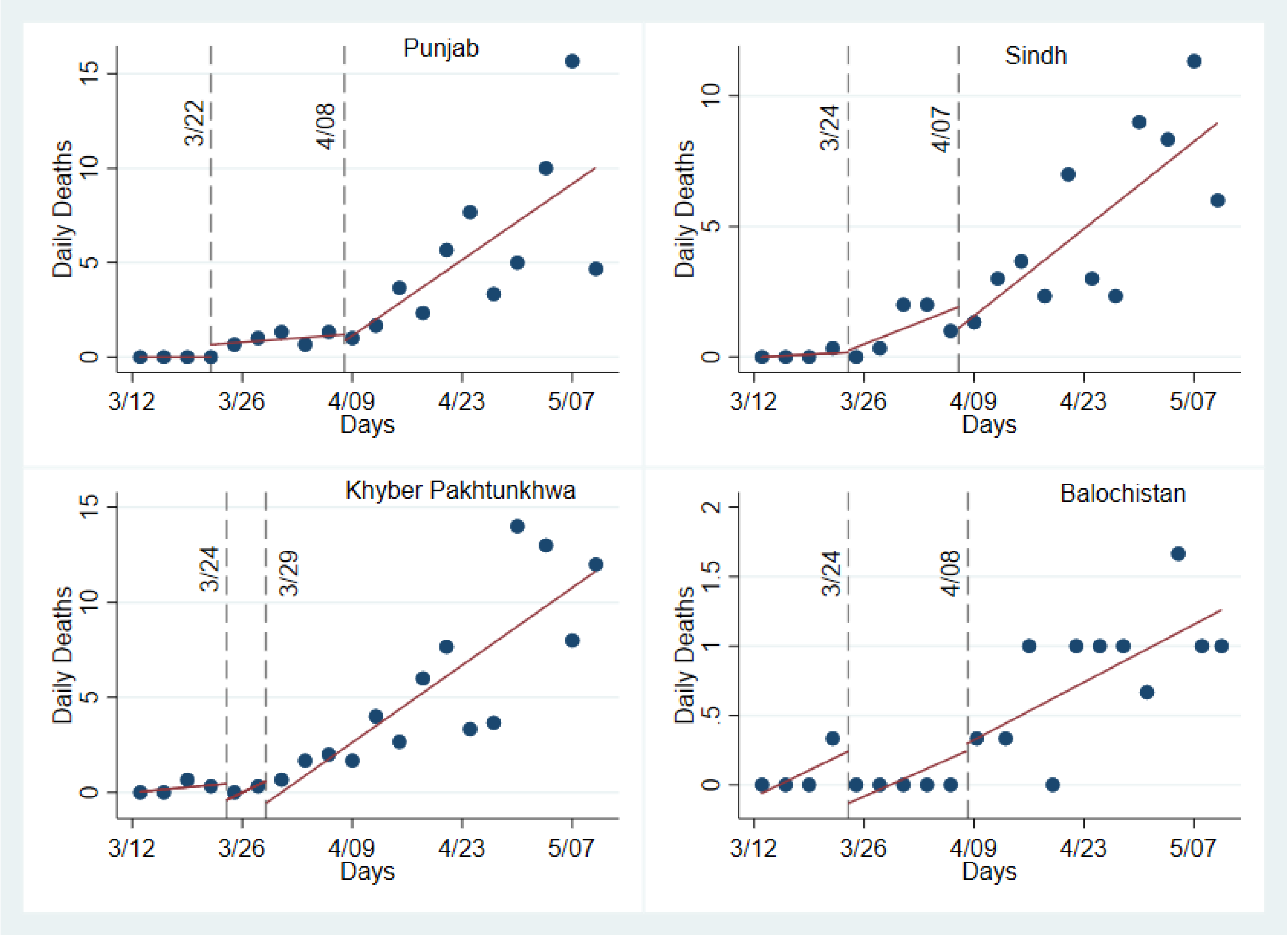

Note: The leftmost vertical dotted line indicates the introduction of a complete lockdown. The rightmost vertical line indicates the introduction of partial lockdown.

Deaths due to Covid-19 increase in Punjab following the implementation of complete lockdown and then increase towards the end of complete lockdown. Following the implementation of the partial lockdown, the deaths initially decrease in Punjab, however, they increase sharply with the passage of time. In Sindh and KPK regions, following complete lockdown deaths initially show a slight decrease; however, they increase towards the end of complete lockdown. Following a partial lockdown, deaths increase in both Sindh and KPK and continue to rise along with time. Overall an increasing trend in deaths is observed during the complete as well as partial lockdown phases in both regions. Following complete lockdown in Balochistan, deaths decrease to zero and remain zero throughout the complete lockdown phase. Following partial lockdown deaths increase and keep on increasing with time,showing a positive trend. Complete lockdown is effective in the containment of deaths in Balochistan only while partial lockdown is ineffective in all four regions.

### 4.2.Poisson and Negative Binomial Regressions

The binned scatter plots reported above indicate the lockdown policies induced heterogeneous regional effects on Covid-19 outcomes. We further analyze the impact of lockdown policies on Covid-19 outcomes systematically by using different regression techniques and estimate the following two equations.

$$\boldsymbol{Daily\ Deaths} = Constant + \alpha \cdot Days + \beta_i \cdot \sum_{i=1}^{7} Complete\ Lockdown + \gamma_j \cdot \sum_{j=1}^{7} Partial\ Lockdown + \delta \cdot Daily\ Positive\ Cases + Daily\ Tests + u \quad \text{-------- (1)}$$

$$\boldsymbol{Daily\ Positive\ Cases} = Constant + \alpha \cdot Days + \beta_i \cdot \sum_{i=1}^{7} Complete\ Lockdown + \gamma_j \cdot \sum_{j=1}^{7} Partial\ Lockdown + Daily\ Tests + u \quad \text{--------- (2)}$$

In the above equations $\sum_{i=1}^{7} Complete\ Lockdown$ represents seven dummy variables for complete lockdown while $\sum_{j=1}^{7} Partial\ Lockdown$ represents seven dummy variables for partial lockdown for seven regions of Pakistan. As complete lockdown and partial lockdown dates are not identical across all seven regions, we add all dummies in the regression equation.

The Covid-19 outcomes (daily deaths and daily positive cases) are count variables; Poisson regression and Negative Binomial Regressions are most relevant for estimating the two aforementioned equations. We first use Poisson regression with robust standard errors to estimate the two equations. The post estimation Pearson Goodness of Fit for both equations indicates the Poisson Model is not the right fit (regression and test results are reported in appendix C). Subsequently, both equations are estimated by using the Negative Binomial Regression with robust standard errors[14], the value of alpha is significant for both equations[15] and indicates Poisson regressions had overdispersion (conditional variance exceeds conditional mean). Hence, we use the output from Negative Binomial regressions for analyzing the effects of lockdown policies in Table 4.

---

[14] The number of observations for each region is not too large and resultantly the panel Negative Binomial Regression (xtnbreg command in STATA 16) estimating equation 1 does not gives results even after large number of iterations. We also tried for clustering the robust standard errors at regional level; however the Wald Chi2 statistic for equation 1 and 2 is not reported by STATA. One of the important reasons for such an outcome is lack of sufficient number of observations. Due to these aforementioned reasons we use simple Negative Binomial Regressions.

[15] LR test is used for the evaluation of alpha. Null hypothesis alpha =0 is rejected at 1% for both equations.

**Table 4: Effects of Lockdown Policies on Covid-19 Outcomes**

| VARIABLES | (1)<br>Daily Deaths | (2)<br>Daily Deaths | (3)<br>Daily Positive Cases | (4)<br>Daily Positive Cases |
|---|---|---|---|---|
| Daily Positive Cases | 0.000768 | 0.000719 | | |
| | (0.000618) | (0.000498) | | |
| Daily Tests Performed | 3.13e-06 | | 0.000171** | |
| | (5.75e-05) | | (8.27e-05) | |
| Days | 0.0446*** | 0.0465*** | 0.0582*** | 0.0655*** |
| | (0.00683) | (0.00674) | (0.00689) | (0.00462) |
| Complete Lockdown ICT | -1.313 | -1.306 | -1.410*** | -1.510*** |
| | (1.112) | (1.112) | (0.347) | (0.330) |
| Partial Lockdown ICT | -0.867 | -0.906 | -1.730*** | -1.948*** |
| | (0.732) | (0.731) | (0.399) | (0.347) |
| Complete Lockdown Punjab | 1.710*** | 1.903*** | 1.583*** | 1.596*** |
| | (0.594) | (0.587) | (0.334) | (0.319) |
| Partial Lockdown Punjab | 2.319*** | 2.312*** | 0.667* | 0.959*** |
| | (0.568) | (0.565) | (0.352) | (0.332) |
| Complete Lockdown Sindh | 2.046*** | 2.065*** | 0.721* | 0.707** |
| | (0.587) | (0.588) | (0.368) | (0.352) |
| Partial Lockdown Sindh | 2.265*** | 2.258*** | 0.790** | 0.962*** |
| | (0.567) | (0.566) | (0.335) | (0.336) |
| Complete Lockdown KPK | -15.65*** | -14.21*** | 0.749* | 0.695* |
| | (0.685) | (0.685) | (0.426) | (0.414) |
| Partial Lockdown KPK | 2.672*** | 2.638*** | 0.242 | 0.117 |
| | (0.570) | (0.569) | (0.356) | (0.319) |
| Complete Lockdown Balochistan | -15.86*** | -14.43*** | -1.071** | -1.167*** |
| | (0.593) | (0.589) | (0.426) | (0.398) |
| Partial Lockdown Balochistan | 0.636 | 0.604 | -0.612 | -0.853** |
| | (0.629) | (0.629) | (0.408) | (0.351) |
| Complete Lockdown AJK | -16.15*** | -14.69*** | -2.890*** | -3.063*** |
| | (0.583) | (0.584) | (0.401) | (0.373) |

| VARIABLES | (1)<br>Daily<br>Deaths | (2)<br>Daily<br>Deaths | (3)<br>Daily<br>Positive<br>Cases | (4)<br>Daily<br>Positive<br>Cases |
|---|---|---|---|---|
| Partial Lockdown AJK | -17.32*** | -15.95*** | -4.166*** | -4.510*** |
| | (0.674) | (0.672) | (0.547) | (0.482) |
| Complete Lockdown GB | -0.514 | -0.565 | -1.406*** | -1.267*** |
| | (0.811) | (0.810) | (0.394) | (0.380) |
| Partial Lockdown GB | -2.308** | -2.354** | -2.635*** | -2.952*** |
| | (1.120) | (1.119) | (0.452) | (0.382) |
| Constant | -982.7*** | -1,025*** | -1,277*** | -1,438*** |
| | (150.1) | (148.2) | (151.4) | (101.6) |
| Observations | 406 | 415 | 407 | 416 |

Robust standard errors are in parentheses. ***p<0.1; **p<0.5; *p<0.1

## 5. Results

Regressions 1 and 2 have daily deaths as the dependent variable, while regressions 3 and 4 have daily positive cases as the dependent variable. In regression 2 and 4, we do not include the explanatory variable for daily tests performed. The exclusion of this variable does not change the significance of other variables in regression 2, while it changes the significance of only one variable in regression 4. Hence, overall the regression results are robust.

### 5.1. Effect of Lockdown Policies on Daily Deaths

While the effect of daily positive cases and daily tests performed on the number of daily deaths is insignificant (regression 1 & 2), the effect of time (variable day) is positive and significant. Holding all other variables constant, with every passing day, the difference in logs of expected counts increases by 0.045 units (regression 1) and by 0.046 units (regression 2). This indicates that the number of deaths in the country is increasing over time. Next, we focus on the main explanatory variables, lockdown and partial lockdown.

In comparison to no lockdown, the effect of complete and partial lockdown on daily deaths is insignificant for the ICT region. Holding all other variables constant, in comparison to no lockdown, complete lockdown has a significant and positive effect on daily deaths for Punjab (difference in logs of expected counts increases by 1.71 in regression 1 and 1.90 in regression 2)[16] and Sindh province, indicating the daily deaths increased during the complete lockdown phase. On the other hand, the complete lockdown variable is significant and negative for KPK (difference in logs of expected counts decreases by 15.65 in regression 1 and 14.21 in regression 2), Balochistan and AJK, indicating daily deaths decreased during the complete lockdown phase. The effect of complete lockdown on daily deaths is insignificant for the GB region. Holding all other variables constant, in comparison to no lockdown, partial lockdown had a significant and positive effect in Punjab, Sindh, and KPK regions. Holding all other variables constant, in comparison to no lockdown, partial lockdown had an insignificant effect on daily deaths in Balochistan, had a significant negative effect in AJK and GB region.

### 5.2.Effect of Lockdown Policies on Daily Positive Cases

Holding all other variables constant, the effect of daily tests performed is significant and positive on daily positive cases, however, the magnitude is very small (difference in logs of expected counts increases by 0.0002 in regression 1). Next, we turn to main explanatory variables. In comparison to no lockdown, complete lockdown had a significant and negative effect on daily positive cases in ICT (difference in logs of expected counts decreases by 1.41 in regression 3 and 1.51 in regression 4), in Balochistan, and GB region. In the three densely populated regions of Punjab, Sindh, and KPK, complete lockdown did not help in preventing the spread of the disease. In comparison to no lockdown, complete lockdown had significant and positive effect on daily positive cases in Punjab (difference in logs of expected counts increases by 1.58 in regression 3 and 1.60 in regression 4), Sindh and KPK. The effect of partial lockdown on daily positive cases is quite similar to complete lockdown. In comparison to no lockdown, the number of positive cases during the partial lockdown decreased in ICT, Balochistan, AJK and GB regions, increased in Punjab and Sindh while did not change significantly in the KPK region.

The effect of complete and partial lockdown on daily deaths, as well as daily positive cases, is heterogeneous across regions. Apparently, thickly populated regions of Punjab, Sindh, and KPK have witnessed an increase in the outcome variables irrespective of the lockdown type; ICT, Balochistan, and GB have shown mixed results while AJK is the only region where both partial and complete lockdowns have effectively reduced the spread of virus and deaths. In the discussion section 7, we explore the possible reasons leading to regional heterogeneity in the outcomes of lockdown policies.

---

[16] We interpret one significant positive and one significant negative coefficient from all four regressions reported in table 4 in an attempt to save space. Rest of the positive and negative coefficients can be interpreted in the identical manner.

## 6. Country Level Analysis – Using Stringency Score as Lockdown Proxy

In this section we turn to another proxy for lockdown- the daily stringency score estimated by Hale et al. (2020). The score varies from zero to 100; a higher value indicates more stringent controls to contain the virus spread. The stringency score is based on the cumulative value of restrictions imposed on schools, workplaces, public events, social gatherings, public transport, stay home orders, domestic and international travel, public information campaign, testing policy, and contact tracing (Hale et al., 2020). Essentially, lockdown and stringency measures are two different ways of quantifying restrictions in an economy.

The stringency score by Hale et al. (2020) is for the whole country, and resultantly, we are unable to do regional analysis using this score. Therefore, we pool the regional data to obtain country-level data and subsequently perform analysis using daily stringency score as the main explanatory variable. The stringency score has 55 observations; its value ranges from 34 to 97, with a mean value of 86.29 and a standard deviation of 18.51. Using the stringency score following two equations are estimated.

$$\boldsymbol{Daily\ Deaths} = Constant + \alpha \cdot Days + \delta \cdot Daily\ Positive\ Cases + Daily\ Tests + Daily\ Stringency\ Score + u \quad \text{----------- (3)}$$

$$\boldsymbol{Daily\ Positive\ Cases} = Constant + \alpha \cdot Days + Daily\ Tests + Daily\ Stringency\ Score + u \quad \text{----------- (4)}$$

Equations (3) and (4) are first estimated by Poisson regression. The Pearson goodness of fit indicates the underlying distribution is not Poisson (results reported in appendix D). Resultantly

Negative Binomial regression is used; the value of alpha is significant[17], indicating Poisson regression had over-dispersion. The regression output is reported in table 5.

**Table 5: Effects of Daily Stringency Score on Covid-19 Outcomes**

| | (1) | (2) | (3) | (4) |
|---|---|---|---|---|
| VARIABLES | Daily Deaths | Daily Deaths | Daily Positive | Daily Positive |
| Daily Positive Cases | 0.000419 | 0.000650* | | |
| | (0.000370) | (0.000355) | | |
| Daily Tests | 3.03e-05 | | 6.65e-05 | |
| | (2.79e-05) | | (4.85e-05) | |
| Days | 0.0493*** | 0.0499*** | 0.0435*** | 0.0610*** |
| | (0.0116) | (0.0115) | (0.0146) | (0.00412) |
| Daily Stringency Score | 0.0362*** | 0.0372*** | 0.00938 | 0.00621 |
| | (0.0136) | (0.0137) | (0.00847) | (0.00790) |
| Constant | -1,087*** | -1,100*** | -952.5*** | -1,338*** |
| | (254.7) | (253.7) | (320.7) | (90.65) |
| Observations | 55 | 55 | 55 | 55 |

In regression 2 and 4, the variable for daily tests is excluded to check the robustness of the effect of the stringency score on the two outcome variables. This exclusion, however, does not alter the significance of the stringency variable.

The effect of daily tests on daily deaths is insignificant in the main regression (1) while it is marginally significant and positive in regression 2; however, the effect is very small. The variable representing days is significant and positive in all four regressions. Holding all other variables constant, with every passing day, the difference in logs of expected counts increases by 0.049 units (regression 1,) by 0.050 units (regression 2), by 0044 (regression 3) and by 0.061 (regression 4). Holding all other variables constant, the effect of daily stringency score is significant and positive on daily deaths (with one unit increase in the stringency score the difference in logs of expected counts increases by 0.036 units in regression 1 and by 0.037 units in regression 2) while it has no significant effect on daily positive cases. These country-level results show that stringency measures appear to be ineffective in the control of damage caused by Covid-19 and in its spread as well.

[17] LR test is used for evaluation of alpha. Null hypothesis alpha =0 is rejected at 1% for both equations.

The regional results show heterogeneity in the effectiveness of lockdown measures while country-level results point towards the ineffectiveness of stringency measures. To reconcile the regional results with country-level results, we need to focus on the three important regions of Punjab, Sindh, and KPK. Together these three regions account for 90% of the total positive cases and 96% of the total deaths during the time period considered in this study. As these three regions represent a huge share of Covid-19 outcomes, they are playing a significant role in driving the country-level statistics. Resultantly, the ineffectiveness of lockdown policies in these regions is reflected in the ineffectiveness of stringency measures at the country-level.

## 7. Discussions and Policy Implications

As discussed earlier, the effect of regional lockdown policies on Covid-19 outcomes is heterogeneous, with very few regions effectively using the lockdown policies to contain the spread of the disease. From a policy perspective, we identify here the factors that have contributed jointly to the ineffectiveness of lockdowns, specifically in the three largest regions of Punjab, Sindh, and KPK.

### 7.1.Confused and Divided Response of Government

The central government in Pakistan has never been unified over the imposition of lockdown measures. The country’s Prime Minister (PM) adamantly opposed lockdown fearing economic impacts on daily wagers that comprise a significant proportion of the country’s labor force. On the other hand, several ministers in the PM’s cabinet proposed strict lockdown measures to contain the virus. As a result of this confusion prevalent in the government circles, the potential risks of Covid-19 were downplayed, the public did not observe the lockdown restrictions seriously, and resultantly the lockdown proved to be ineffective in three large provinces[18].

[18] Sources:
https://www.brookings.edu/blog/order-from-chaos/2020/03/27/pakistan-teeters-on-the-edge-of-potential-disaster-with-the-coronavirus/
https://thediplomat.com/2020/05/pakistan-plans-another-covid-19-lockdown-will-it-work/
https://www.voanews.com/covid-19-pandemic/lockdown-or-no-lockdown-confusion-dominates-pakistans-covid-response
https://thewire.in/health/in-south-asia-lanka-leads-and-india-lags-in-infrastructure-medical-response-to-covid-19

### 7.2. Poor Habitat Conditions

A significant proportion of Pakistan's population inhibits in Punjab (110 million), Sindh (47.9 million) and KPK (30.5 million) regions (Wazir & Goujon, 2019). Together these three regions account for approximately 89% of the country's population, and a major proportion of this population resides in rural areas or slums in large cities (for example Sindh's capital city Karachi has the world's largest slum population approximated to be 2 million[19]). In rural and specifically slum areas, social distancing or keeping one restricted to home are almost non-existent, issues of cleanliness are acute, and poverty rates are high. All these factors, when combined together, provided an ideal habitat for the sharp spread of Covid-19, even in the presence of lockdown measures.

### 7.3. Informal Labor Force

The informal labor force makes up 75% of the total labor force (48.75 million). About 40% of this informal labor force (19.5 million) is employed by the agriculture sector, while 60% (29.25 million) works for the industrial and service sector.[20] The majority of the informal labors are paid on a daily basis and unfortunately forced to leave homes in an attempt to earn money for food and subsequently violate and undermine the effectiveness of lockdown restrictions by serving as a potential source of virus spread within their workplace and residences.

### 7.4. Religious Gatherings

Pakistan is a religiously homogeneous country with Muslims making up 96.28% of the population.[21] In Islam, daily five congregational prayers in the mosque are a vital part of worship. Congregations and gatherings could be favorable grounds for viral transmission. In Pakistan, the religious, as well as political leaders, were divided on the closing down of mosques. Resultantly, government in coordination with religious leaders set up protocols for congregational prayers in mosques, however in most cases these protocols were not strictly followed, possibly leading to rapid transmission of virus even during the lockdown phases.[22]

---

[19]Source: Nikkei Asian Review https://asia.nikkei.com/Spotlight/Coronavirus/India-and-Pakistan-s-red-zones-keep-COVID-19-trajectory-on-rise

[20] Source: https://www.aa.com.tr/en/asia-pacific/millions-of-pakistani-laborers-struggle-amid-covid-19-lockdown/1824231

[21] Source: Pakistan Bureau of Statistics. Accessed at: http://www.pbs.gov.pk/sites/default/files/tables/POPULATION%20BY%20RELIGION.pdf

[22] Sources: VOA and Aljazeera News.

### 7.5. Public Behavior

As an entirely unprecedented situation for masses, the heterogeneous behavioral response of the public to lockdown was evident. In the case of Pakistan, poor handling of the situation by the government over using force to implement lockdown strategies in the big provinces led to fear in the minds and created a social stigma around Covid-19. Subsequently, people with symptoms kept on living with family members, and whole families got infected. These numbers eventually started showing up post lockdown in the largest regions. A lack of literacy and misinformation about the disease and its treatment further complicated the outcomes. On the contrary, in AJK and GB, where people were willing to observe lockdown, showed better outcomes. One reason why they showed this willingness is likely their high literacy rate than the national average rate.

From a policy perspective, it is clear that lockdown effectively used by rich countries, including Germany, Japan, and USA, is to a large extent, ineffective in controlling the spread of virus in a poor country like Pakistan. In fact, several researchers have already discussed the possibility of the lockdown being ineffective in poor countries (for example Barnett-Howell & Mobarak, 2020; Cash & Patel, 2020). The country-level results from this article support the predictions of these studies. We think contextualized strategies would be way more effective in the control of the virus in poor countries. In the case of Pakistan, the socio-economic and political conditions and religious norms are some of the important contextual elements that should be considered while making future lockdown strategies.

## Conclusion:

The purpose of the paper was to examine the impact of lockdown strategies in different regions of Pakistan. The three large regions of Punjab, Sindh, and KPK have witnessed an increase in positive cases as well as deaths despite the type of lockdown measures. The results for ICT, Balochistan, and GB regions are mixed depending on the type of lockdown. AJK is the only region that has been successful in the control of virus with both complete and partial lockdowns. The country-level results indicate that stringency measures have been ineffective in the containment of the virus. The regional differential in Covid-19 outcomes observed in the

---

https://www.voanews.com/covid-19-pandemic/lockdown-or-no-lockdown-confusion-dominates-pakistans-covid-response

https://www.aljazeera.com/news/2020/05/pakistan-mosques-coronavirus-battleground-issue-200504102030000.html

case of Pakistan suggests similar lockdown strategies can have different effects within a single country. In Pakistan divided political leadership, weak public will, and lockdown enforcement mechanisms, socio-economic constraints, and epidemic stage have collectively influenced the ineffectiveness of lockdown policies. Now when the government has eased out lockdown restrictions and there is a huge surge in the caseload, the country is likely going towards a herd immunity model. Before the situation goes out of control, the government can still contain the spread of the disease by taking following actions:

- Federal government (particularly, National Command and Operation Center, which devises and implements strategies about Covid-19) should facilitate the provinces by ramping up their testing capabilities, enforcing isolation, and implementing contact tracing strategies.
- The local governments should proactively work together to devise and implement comprehensive strategies to stop the spread of the disease.
- Federal and local governments should involve local religious scholars and community elders in teaching the public the importance of social distancing, hygiene, and prevention. They should also work on dispelling and addressing the stigma around Covid-19 so that people can test themselves without anxiety and fear from society.
- Government should allow shorter working hours, limit number of people at religious and social gatherings, and implement universal masking using cloth masks for the community.

## Appendix A: Lockdown Policies & Spread of Covid-19

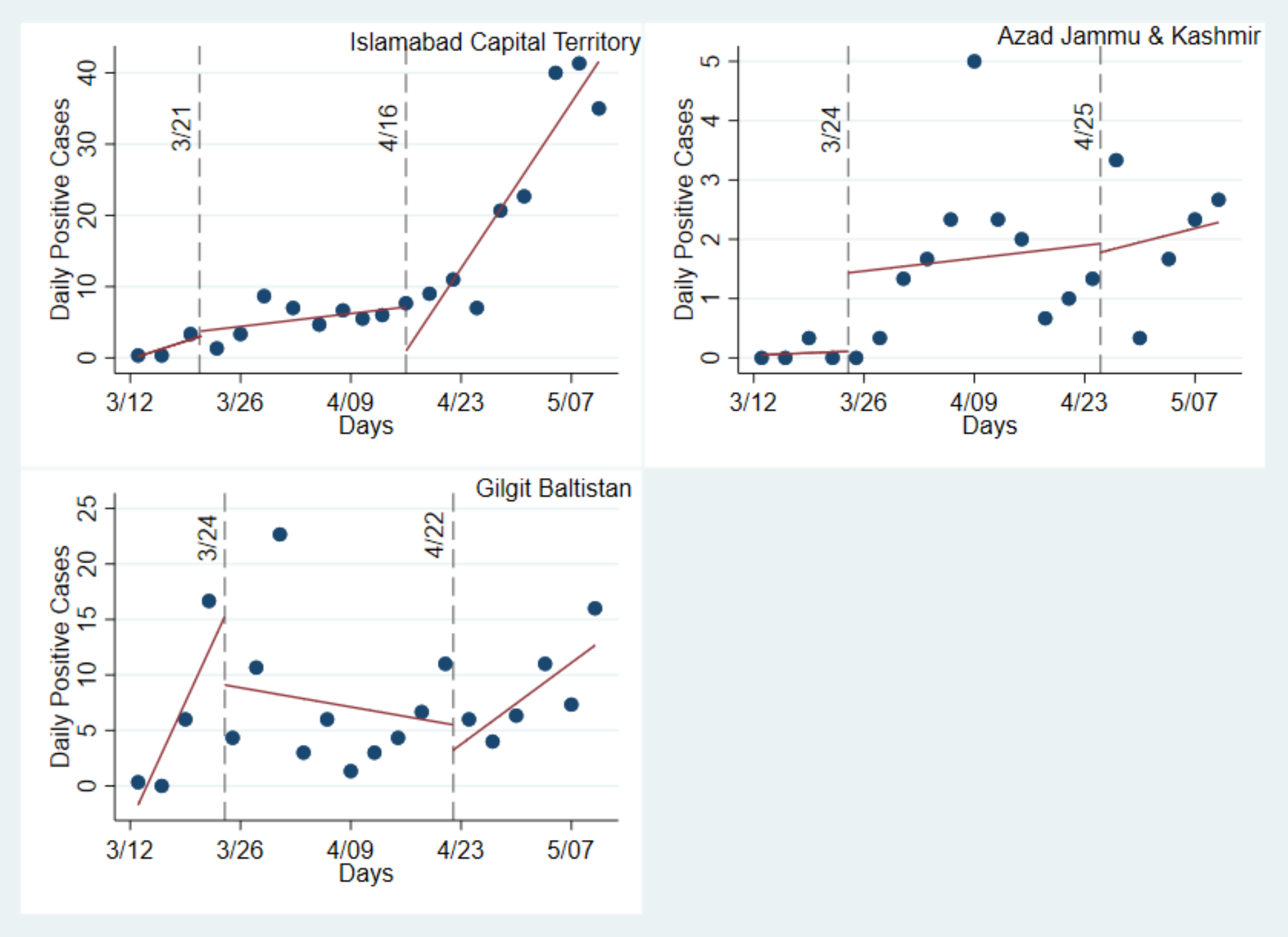

Note: The leftmost vertical dotted line indicates the introduction of a complete lockdown. The rightmost vertical line indicates the introduction of partial lockdown.

## Appendix B: Lockdown Policies & Deaths due to Covid-19

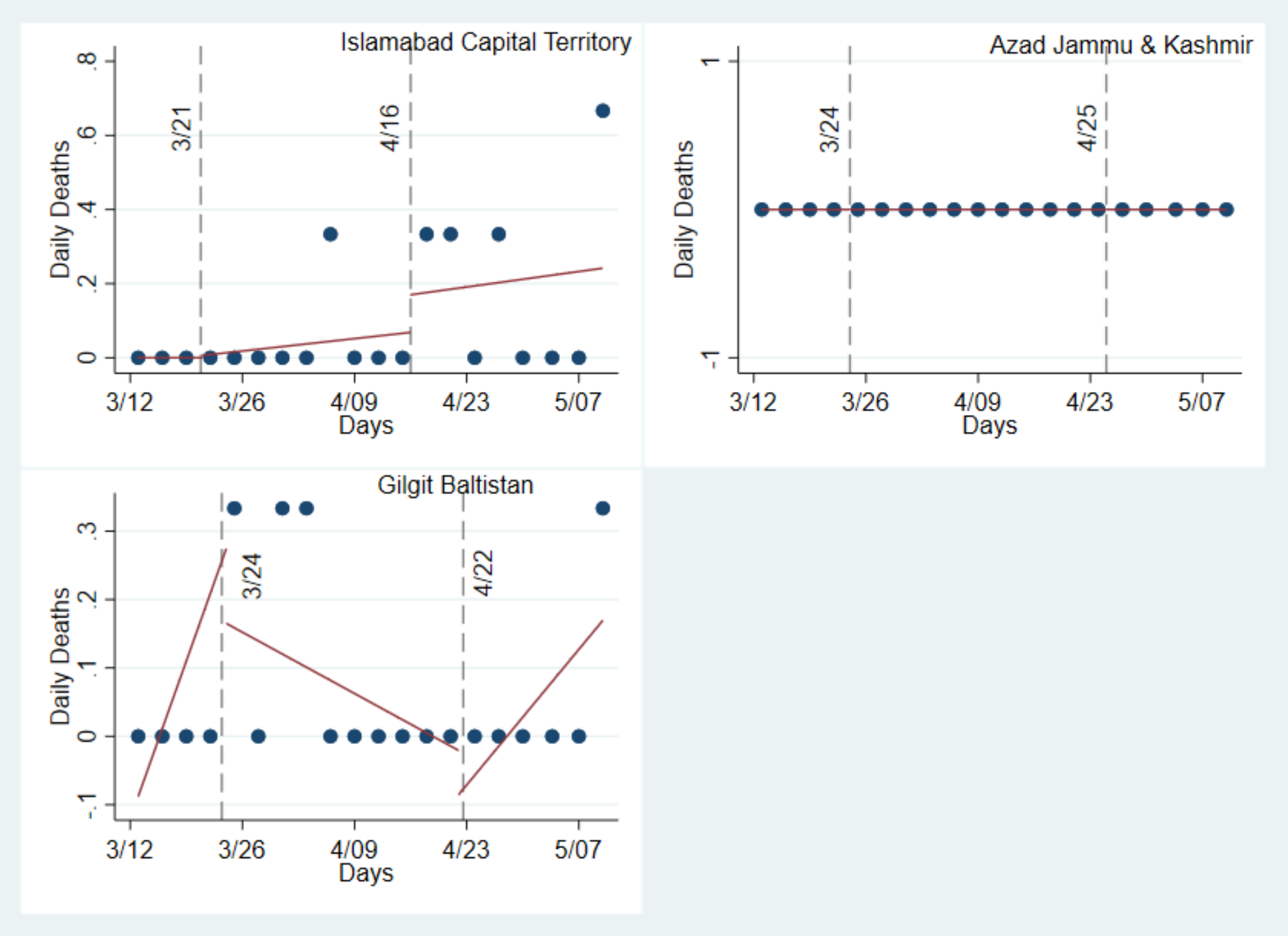

Note: The leftmost vertical dotted line indicates the introduction of a complete lockdown. The rightmost vertical line indicates the introduction of partial lockdown.

**Appendix C: Results of Poisson Regression (Lockdown Main IV)**

| VARIABLES | (1)<br>Daily Deaths | (2)<br>Daily Deaths | (3)<br>Daily Positive Cases | (4)<br>Daily Positive Cases |
|---|---|---|---|---|
| Daily Positive Cases | 0.000345 | 0.000497 | | |
| | (0.000526) | (0.000413) | | |
| Daily Tests Performed | 4.00e-05 | | 0.000111*** | |
| | (5.58e-05) | | (3.84e-05) | |
| Days | 0.0438*** | 0.0455*** | 0.0537*** | 0.0632*** |
| | (0.00623) | (0.00624) | (0.00501) | (0.00180) |
| Lockdown ICT | -1.306 | -1.292 | -1.499*** | -1.670*** |
| | (1.111) | (1.111) | (0.338) | (0.526) |
| Partial Lockdown ICT | -0.851 | -0.865 | -1.548*** | -1.907*** |
| | (0.714) | (0.712) | (0.374) | (0.563) |
| Lockdown Punjab | 1.712*** | 1.938*** | 1.533*** | 1.617*** |
| | (0.606) | (0.587) | (0.381) | (0.518) |
| Partial Lockdown Punjab | 2.360*** | 2.421*** | 0.863** | 0.883 |
| | (0.563) | (0.557) | (0.351) | (0.561) |
| Lockdown Sindh | 2.059*** | 2.092*** | 0.667* | 0.593 |
| | (0.586) | (0.585) | (0.359) | (0.518) |
| Partial Lockdown Sindh | 2.317*** | 2.345*** | 1.027*** | 0.968* |
| | (0.563) | (0.561) | (0.348) | (0.560) |
| Lockdown KPK | -14.79*** | -17.35*** | 0.703* | 0.633 |
| | (0.684) | (0.684) | (0.420) | (0.510) |
| Partial Lockdown KPK | 2.748*** | 2.724*** | 0.346 | 0.0548 |
| | (0.569) | (0.567) | (0.355) | (0.558) |
| Lockdown Balochistan | -14.97*** | -17.53*** | -1.147*** | -1.275** |
| | (0.590) | (0.586) | (0.382) | (0.519) |
| Partial Lockdown Balochistan | 0.671 | 0.644 | -0.418 | -0.788 |
| | (0.619) | (0.620) | (0.371) | (0.561) |
| Lockdown AJK | -15.23*** | -17.76*** | -2.997*** | -3.253*** |
| | (0.574) | (0.575) | (0.394) | (0.537) |
| Partial Lockdown AJK | -16.40*** | -19.02*** | -4.094*** | -4.550*** |
| | (0.653) | (0.654) | (0.496) | (0.567) |
| Lockdown GB | -0.509 | -0.556 | -1.660*** | -1.673*** |
| | (0.794) | (0.793) | (0.372) | (0.533) |
| Partial Lockdown GB | -2.273** | -2.309** | -2.555*** | -2.993*** |

| VARIABLES | (1) Daily Deaths | (2) Daily Deaths | (3) Daily Positive Cases | (4) Daily Positive Cases |
|---|---|---|---|---|
| | (1.118) | (1.118) | (0.416) | (0.566) |
| Constant | -965.2*** | -1,004*** | -1,178*** | -1,389*** |
| | (137.1) | (137.3) | (110.2) | (39.16) |
| Observations | 406 | 415 | 407 | 416 |
| Pearson goodness-of-fit | 464.0136 | 477.3982 | 10334.89 | 11525.76 |
| Prob > chi2 | 0.0048 | 0.0038 | 0 | 0 |

Robust standard errors are in parentheses. ***p<0.1; **p<0.5; *p<0.1

**Appendix D: Results of Poisson Regression (Stringency Score Main IV)**

| VARIABLES | (1) Daily Deaths | (2) Daily Deaths | (3) Daily Positive Cases | (4) Daily Positive Cases |
|---|---|---|---|---|
| Daily Positive Cases | 0.000242 | 0.000568*** | | |
| | (0.000280) | (0.000217) | | |
| Daily Tests Performed | 3.99e-05 | | 7.27e-05*** | |
| | (2.80e-05) | | (1.95e-05) | |
| Days | 0.0495*** | 0.0486*** | 0.0395*** | 0.0590*** |
| | (0.00901) | (0.00944) | (0.00723) | (0.00431) |
| Stringency Score | 0.0343*** | 0.0361*** | 0.00708 | 0.00366 |
| | (0.0120) | (0.0121) | (0.00763) | (0.00749) |
| Constant | -1,091*** | -1,072*** | -864.2*** | -1,293*** |
| | (198.2) | (207.6) | (158.8) | (94.69) |
| Observations | 55 | 55 | 55 | 55 |
| Pearson goodness-of-fit | 83.14676 | 85.7571 | 2761.469 | 4037.822 |
| Prob > chi2 | 0.0022 | 0.0017 | 0 | 0 |

Robust standard errors are in parentheses. ***p<0.1; **p<0.5; *p<0.1